\documentclass[fleqn]{mnras}
\usepackage{graphicx}	
\usepackage{amsmath}	
\usepackage{amssymb}	

\title[Nonthermal Pressure in A2142]{Nonthermal Pressure in the Outskirts of Abell 2142}

\author[Fusco-Femiano \& Lapi]{Roberto Fusco-Femiano$^{1}$\thanks{E-mail: roberto.fuscofemiano@iaps.inaf.it} and Andrea Lapi$^{2,3,4}$
\\
$^{1}$IAPS-INAF, Via Fosso del Cavaliere, 00133 Roma,
Italy\\
$^{2}$SISSA, Via Bonomea 265, 34136 Trieste, Italy\\
$^{3}$INAF-Osservatorio Astronomico di Trieste, via Tiepolo 11, 34131 Trieste, Italy\\
$^{4}$INFN-Sezione di Trieste, via Valerio 2, 34127 Trieste, Italy
}

\date{Accepted XXX. Received YYY; in original form ZZZ}

\pubyear{2017}

\begin{document}
\label{firstpage}
\pagerange{\pageref{firstpage}--\pageref{lastpage}}
\maketitle

\begin{abstract}
Clumping and turbulence are expected to affect the matter accreted onto the outskirts of galaxy clusters. To determine their impact on the thermodynamic properties of Abell 2142 we perform an analysis of the X-ray temperature data from \textit{XMM-Newton} via our SuperModel, a state-of-the-art tool for investigating the astrophysics of the intracluster medium already tested on many individual clusters (since Cavaliere et al. 2009). Using the gas density profile corrected for clumpiness derived by Tchernin et al. (2016), we find evidence for the presence of a nonthermal pressure component required to sustain gravity in the cluster outskirts of Abell 2142, that amounts to about $30\%$ of the total pressure at the virial radius. The presence of the nonthermal component implies the gas fraction to be consistent with the universal value at the virial radius and the electron thermal pressure profile to be in good agreement with that inferred from the SZ data. Our results indicate that the presence of gas clumping and of a nonthermal pressure component are both necessary to recover the observed physical properties in the cluster outskirts. Moreover, we stress that an alternative method often exploited in the literature (included Abell 2142) to determine the temperature profile $k_BT = P_e/n_e$ basing on a combination of the Sunyaev-Zel'dovich (SZ) pressure $P_e$ and of the X-ray electron density $n_e$ does not allow to highlight the presence of nonthermal pressure support in the cluster outskirts.
\end{abstract}

\begin{keywords}
cosmic background radiation --- galaxies: clusters: individual
(Abell 2142) --- X-rays: galaxies: clusters
\end{keywords}

\section{Introduction}

Cluster outskirts provide the connection between the intracluster medium (ICM) and the filamentary structures of the cosmic web, and contain substantial amounts of the baryons and of the gravitationally dominant dark matter components. These external regions are extremely interesting since they are the sites of several physical processes and events affecting the thermodynamic properties of the intracluster medium (ICM; see Kravtsov \& Borgani 2012; Cavaliere \& Lapi 2013; Reiprich er al. 2013). The matter accreted onto the cluster outskirts is expected to be clumpy as well as affected by nonthermal energy inputs in the form of turbulence, bulk motions, cosmic ray pressure and magnetic fields, as reported by many empirical investigations and hydrodynamical simulations (Vazza et al. 2009; Valdarnini 2011; Lau et al. 2013; Gaspari \& Churazov 2013; Nelson et al. 2014). Deviations from spherical symmetry (e.g., Morandi et al. 2010; Donahue et al. 2014), and the presence of a nonthermal pressure support can contribute to originate the bias between the estimates of the cluster mass from gravitational lensing and from X-ray/Sunyaev-Zel'dovich (SZ) effect under the assumption of strictly thermal hydrostatic equilibrium (Martizzi \& Agrusa 2016).

The advent of the \textsl{Suzaku} X-ray observatory with its low and stable particle background has made possible to spectroscopically study cluster outskirts in X rays. The main, somewhat unexpected, findings from the \textsl{Suzaku} observations can be briefly summarized as follows: the ICM temperature rapidly declines outward in the region $r \sim 0.3 - 1\, r_{200}$ (Akamatsu et al. 2011; Reiprich et al. 2013); the entropy profile flattens at $r \gtrsim 0.5 r_{200}$ (see Walker et al. 2013) relative to the power law increase with slope $1.1$ expected from strongly shocked accretion of external gas under pure gravitational infall (Tozzi \& Norman 2001; Lapi et al. 2005; Voit 2005); the thermodynamic properties of the ICM are subject to significant azimuthal variations, with a more efficient thermalization observed in infall regions facing filamentary structures of the cosmic web (Kawaharada et al. 2010; Ichikawa et al. 2013; Sato et al. 2014); in some relaxed clusters the X-ray mass profile features an unphysical, decreasing behavior at large radii (e.g., Ichikawa et al. 2013; Sato et al. 2014) commonly interpreted in terms of an ICM far from strictly thermal hydrostatic equilibrium (HE); the latter effect in turn implies a gas mass fraction $f_{\rm gas}$ higher than the cosmic value (Simionescu et al. 2011; Fusco-Femiano \& Lapi 2013, 2014).

Several mechanisms have been proposed to explain such observational findings. The entropy flattening may be due to clumping (e.g., Simionescu et al. 2011; Nagai \& Lau 2011; Walker et al. 2013) in the gas density distribution; this causes an overestimate of the ICM density and hence an underestimate of the entropy in the outskirts. Low-entropy gas in the outskirts may be originated because of the different timescales for thermalization of ions and electrons by Coulomb collisions (Hoshino et al. 2010; Akamatsu et al. 2011; Avestruz
et al. 2015). Weakening of the accretion shocks due to slow accretion at late cosmic times or in cluster sectors facing a void (Lapi et al. 2010; Cavaliere et al. 2011) may imply entropy flattening, steep temperature decline, nonthermal motions, and azimuthal variations. The acceleration of cosmic rays by part of the gas infall energy may lead to a weaker entropy generation at the boundary shocks (Fujita et al. 2013). All these mechanisms are described with some more details in Fusco-Femiano \& Lapi (2015).

Recently, Tchernin et al. 2016 (hereafter T16) have reported the thermodynamic properties of the massive galaxy cluster Abell 2142 out to $2\, \times r_{500}$. Adopting the multi-scale method described in Eckert et al. (2016), these authors combine the pressure profile derived from SZ data by \textsl{Planck} and the brightness distribution observed in X rays by \textsl{XMM-Newton} to infer the gas density profile corrected for clumpiness. Their conclusion is that the thermodynamic properties of Abell 2142 are compatible with purely thermal gas pressure in HE, and that the density profiles corrected for clumpiness can explain the entropy flattening observed by \textsl{Suzaku} in the outskirts of several clusters, avoiding in such way the need to consider other mechanisms.

In the present paper we perform the SuperModel (SM; Cavaliere et al. 2009; Fusco-Femiano et al. 2009) analysis of Abell 2142 using the X-ray temperature data from \textsl{XMM-Newton} at $r \lesssim r_{500}$, and adopting the gas density profile corrected for clumpiness by T16. The SM is a state-of-the-art semianalytic tool to investigate the clusters' thermodynamic properties based on minimal physical assumptions on the underlying entropy state of the ICM; it has been successfully applied to the analysis of several individual clusters (Fusco-Femiano et al. 2009, 2011), and more recently exploited to assess the role of nonthermal pressure support in the cluster outskirts (Fusco-Femiano \& Lapi 2013, 2014, 2015). Our analysis for Abell 2142 will reveal the presence of an appreciable nonthermal pressure component, apparently at variance with the SZ/X-ray analysis. In fact, we will also shed light on why the method based on the combination of SZ pressure and X-ray density data cannot highlight such a nonthermal component.

The plan of the paper is straightforward. In Section 2 we analyze with the SM the temperature profile of Abell 2142 as measured out to $r_{500}$ by \textsl{XMM-Newton}. For the fitting procedure we exploit the SM equations reported in Appendix A (see also Fusco-Femiano \& Lapi 2013), yielding the temperature, pressure and total mass when a nonthermal pressure component is included in the HE equation. We discuss the results and draw our conclusions in Section 3.

Throughout the paper we adopt the standard flat cosmology (\textsl{Planck} collaboration XIII 2016) with parameters in round numbers: $H_0 = 70$ km s$^{-1}$ Mpc$^{-1}$, $\Omega_{\Lambda} = 0.7$, $\Omega_M = 0.3$. At the redshift of Abell 2142 ($z = 0.09$; Owers et al. 2011), 1 arcmin corresponds to $102$ kpc. In our fit we have assumed $R = 2\times r_{500} =$ 2816 kpc where $r_{500}$ = 1408 kpc, used in T16, is the result of a joint analysis performed by Munari et al. (2014) for Abell 2142.

\begin{figure}
    \includegraphics[width=\columnwidth]{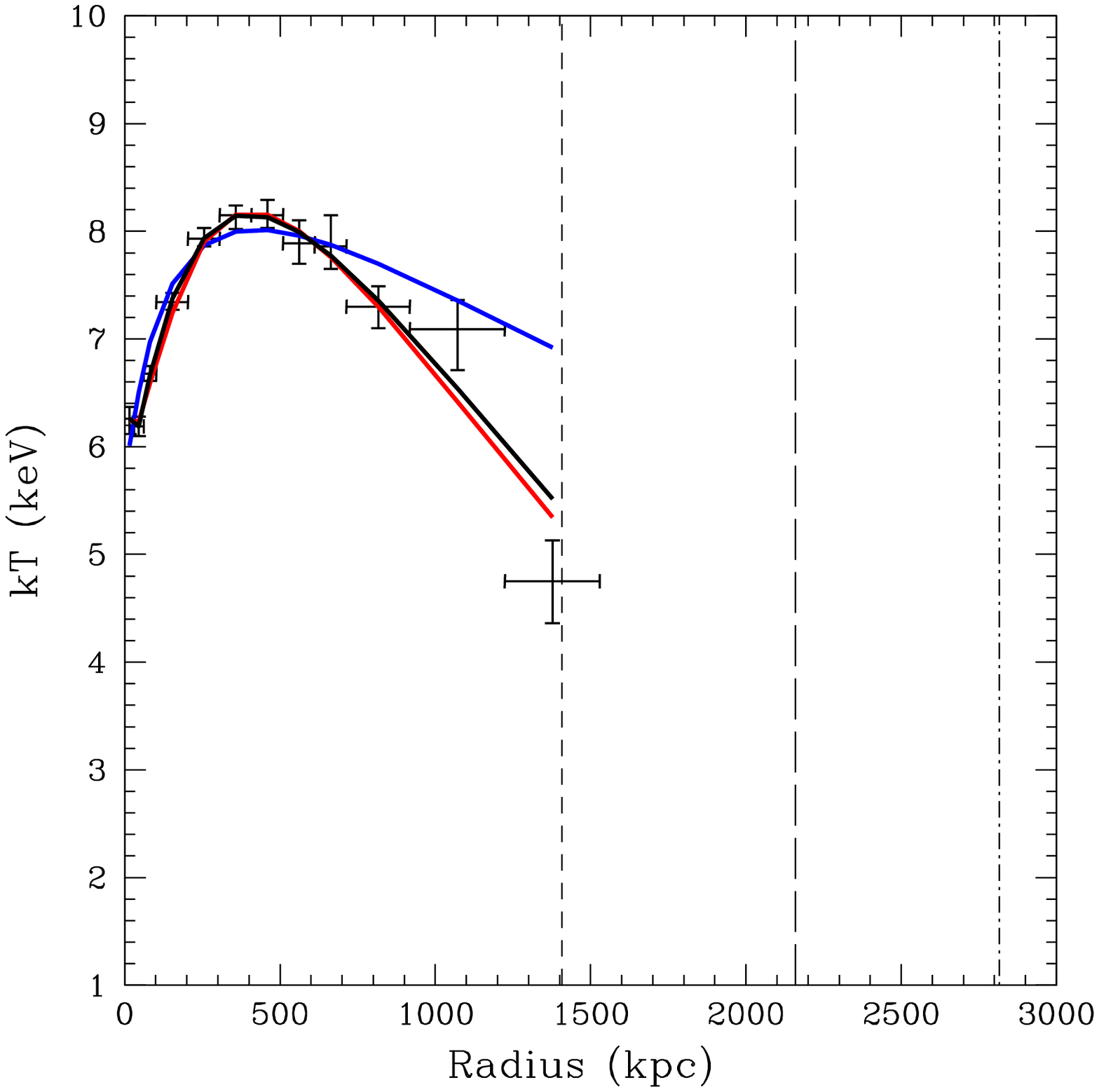}\\\includegraphics[width=\columnwidth]{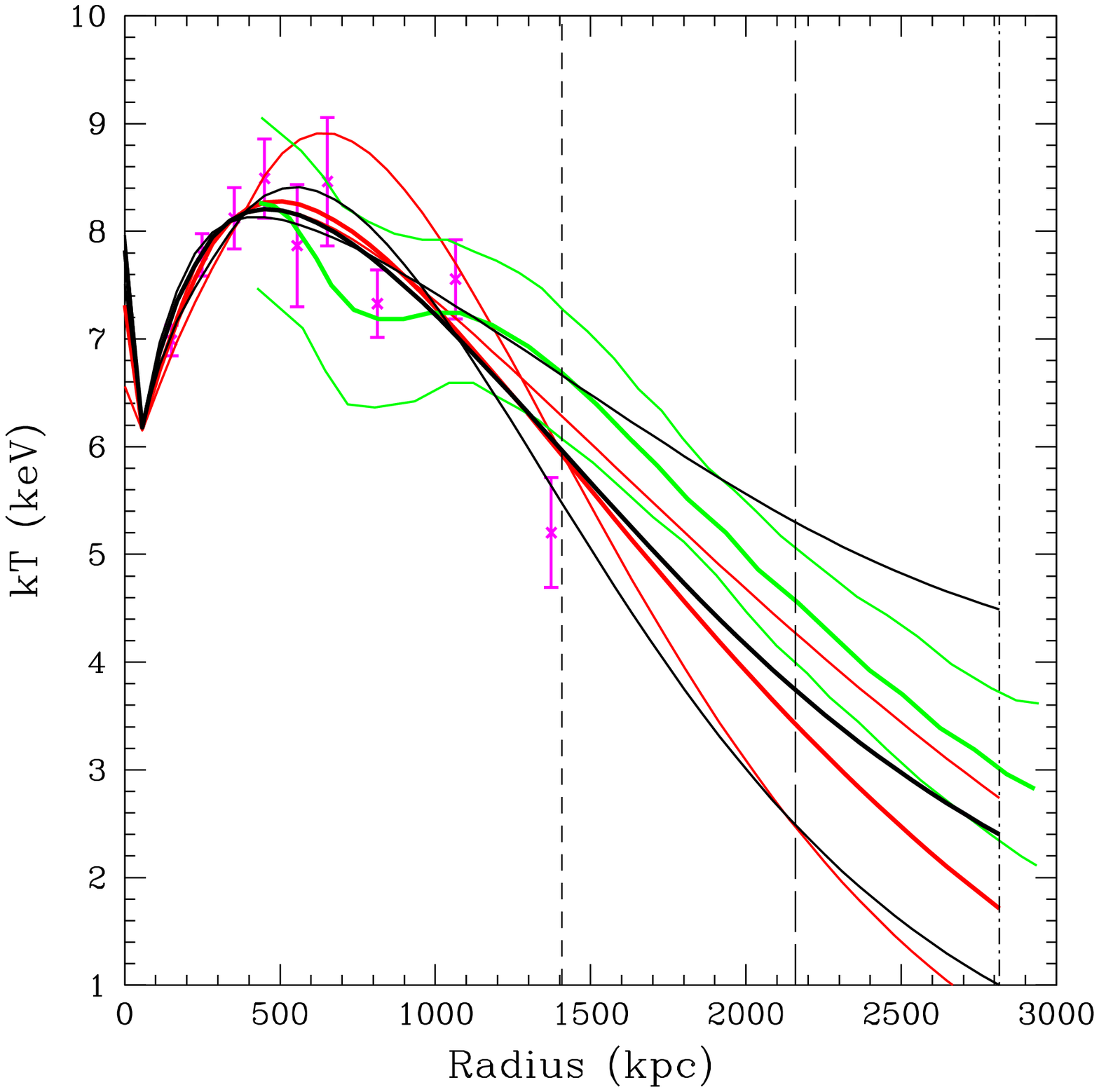}
    \caption{Abell 2142 - Top panel: Projected temperature profiles. Data points from \textsl{XMM-Newton} (Tchernin et al. 2016). Blue and red lines are the SM fit in the absence of turbulence in the cluster outskirts ($\delta_R$ = 0); the black line is the fit with $\delta_R$ = 0.5 and $l$ = 0.4 (see Eq. [1]). The blue line is obtained with a power law increase of the entropy while the red and black curves are derived with an entropy profile that deviates from a power law at $r \gtrsim r_b$ (see text). Bottom panel: Deprojected temperature profiles. The thick red and black lines correspond to the red and black fits reported in the top panel (the thin lines represent the uncertainties at the $1\sigma$ level); the green line is the deprojected temperature profile obtained by T16 combining the SZ pressure profile measured by \textsl{Planck} with the \textsl{XMM-Newton} gas density profile corrected for clumpiness. Data points represent the spectroscopic data (see top panel) deprojected by T16 using the method of Ettori et al. (2010). The vertical dashed, long-dashed and dash-dotted lines represent $r_{500}$, $r_{200}$ and the virial radius $R$, respectively.}
\end{figure}

\section{SuperModel analysis of Abell 2142}

We analyze the \textsl{XMM-Newton} projected temperature profile with our SM, specifying in the HE equation (possibly including a nonthermal component, see below) the entropy profile $k(r) \equiv k_B T/n^{2/3}$. For this analysis we assume a spherically averaged profile with shape $k(r) = k_c + (k_R - k_c)(r/R)^a$ (see Lapi et al. 2005; Voit 2005) where $k_c$ is the central entropy level, set by the balance between various production and erosion processes including feedback from astrophysical sources and radiative cooling (see Cavaliere et al. 2009 for details); $k_R$ is the entropy at the virial radius $R$ produced by supersonic inflows of gas from the external environment into the cluster gravitational potential well; $a$ is the slope of the power law increase from the central entropy value. The above entropy shape has been shown to be consistent with simulations and observations of both stacked and individual clusters (see Voit 2005; Cavagnolo et al. 2009; Fusco-Femiano et al. 2009; Pratt et al. 2010; Fusco-Femiano \& Lapi 2013, 2014, 2015).

We also use a modified entropy run, as it has been considered in several cool-core clusters for taking into account the steep temperature and flat entropy profiles observed by \textsl{Suzaku} toward the virial radius (see Lapi et al. 2010; Walker et al. 2013). The modified shape starts as a simple power law with slope $a$, but for radii $r \gtrsim r_b$ deviates downward; for the sake of simplicity, the entropy slope is taken to decline linearly with a gradient $a^{\prime}\equiv (a - a_R)/(R/r_b -1)$, where $r_b$ and $a^{\prime}$ are free parameters to be determined from the fitting of the X-ray observables. Such a shape is meant to render the effects of a reduced entropy production relative to a pure gravitational inflow; this occurs when the accretion rates gradually decrease and the accretion shocks weaken due to the slowdown at low $z$ of the cosmological structure growth in an accelerating universe (Lapi et al. 2010; Cavaliere et al. 2011; Fusco-Femiano \& Lapi 2014).

At the same time, the weakening of accretion shocks in cluster outskirts is expected to let relatively more bulk energy seep through the cluster and drive turbulence into the ICM (Cavaliere et al. 2011; Fusco-Femiano \& Lapi 2014). In the presence of turbulent motions, an additional nonthermal component contributes to sustain gravity in the HE equation. The total pressure can be written as $p_{\rm tot}(r) = p_{\rm th} + p_{\rm nth} = p_{\rm th}[1 + \delta (r)]$ in terms of the quantity $\delta (r) \equiv p_{\rm nth}/p_{\rm th}$, where $p_{\rm th}$ and $p_{\rm nth}$ are the thermal and nonthermal pressure, respectively. For $\delta(r)$ we use the functional shape that concurs with the indication of numerical simulations (e.g., Lau et al. 2009; Vazza et al. 2011)
\begin{equation}
\delta(r) = \delta_R\, e^{-(R-r)^2/l^2}
\end{equation}
which takes on the maximum value $\delta_R$ at the virial radius $R$ and decays toward the inner cluster regions over the characteristic length-scale $l$.

Along these lines we have analyzed the projected X-ray temperature profile observed by \textsl{XMM-Newton} in Abell 2142 (T16), adopting the two entropy runs discussed above (e.g., a pure powerlaw profile, and a profile bending outwards). All in all, a better fit is obtained for the entropy profile bending outwards, as shown by the red line in Fig. 1 (top panel); note that this fit still does not include a nonthermal pressure component. The corresponding 3D temperature profile (see Fig. 1, bottom panel) is subject to a large error toward the virial radius because the existing data fitted by the SM are limited to $r_{500}$. However, we shall show that our SM mass profiles are consistent with the measurements reported at distances greater than $r_{500}$, highlighting the reliability of the SM temperature fit also toward the cluster outskirts. The SM deprojected temperature profile is consistent at $r\lesssim r_{500}$ with that derived by T16, via combination $k_B\,T = P_e/n_e$ of the deprojected SZ pressure $P_e$ profile (\textsl{Planck} collaboration V 2013) and of the X-ray density profile $n_e$; note that the latter profile has been corrected for clumpiness. Contrariwise, at $r \gtrsim r_{500}$ the SM temperature profile is marginally consistent with the SZ/X-ray outcome.

\begin{figure}
    \includegraphics[width=\columnwidth]{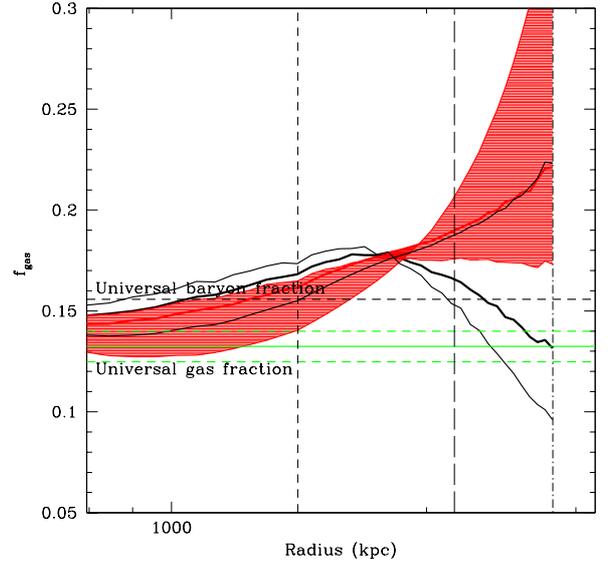}
    \caption{Abell 2142 - Gas fraction profile $f_{\rm gas} = M_{\rm gas}/M$. The thick red profile is obtained with $M_{\rm gas}$ derived by the gas density corrected for clumpiness (see T16) and by the total mass $M$ derived with $\delta_R = 0$ (see Fig. 4, top panel). The hatched area is the uncertainty at the $1\sigma$ level. The black profile is derived with $\delta_R = 0.5$ and $l=0.4$; the thin black lines are the uncertainty at the $1\sigma$ level. The dashed horizontal line represents the universal baryon fraction from \textsl{Planck} (Planck Collaboration XIII 2016), whereas the thick green line with the dashed green lines is the expected gas fraction corrected for the baryon fraction in the form of stars (Gonzalez et al. 2007).}
\end{figure}

\begin{figure}
    \includegraphics[width=\columnwidth]{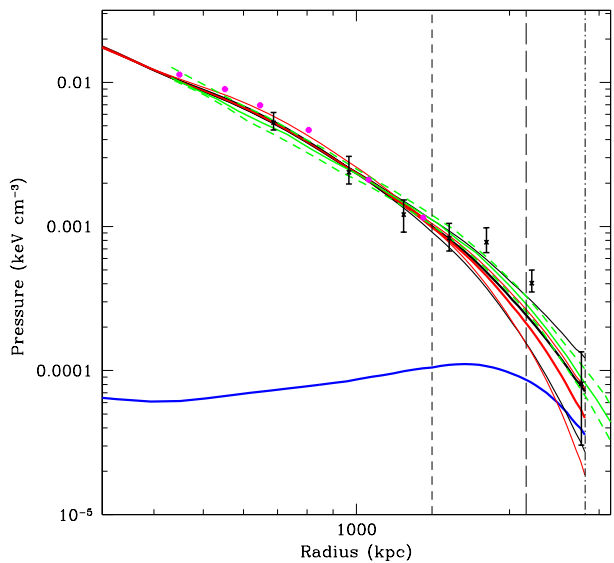}
    \caption{Abell 2142 - Electron pressure profile. The red and black curves represent the thermal pressure derived from the deprojected temperature profiles with $\delta_R = 0$ and $\delta_R = 0.5$ ($l$ = 0.4), respectively (see Fig. 1, bottom panel) and with the density profile corrected for clumpiness (see T16). The blue curve is the nonthermal pressure component (see Eq. [1]) with $\delta_R = 0.5$ and $l$ = 0.4. The green lines and the black points represent the best-fit pressure profiles obtained by T16 deprojecting the SZ data using two methods (see their caption of Fig. 10); the magenta points show the spectroscopic X-ray data deprojected by T16 with the method of Vikhlinin et al. (2006). The vertical dashed, long-dashed and dot-dashed lines represent $r_{500}$, $r_{200}$ and the virial radius $R$, respectively.}
\end{figure}

Combining the 3D temperature profile with the density profile corrected for clumpiness we recover the radial distribution of gas fraction, pressure, hydrostatic mass and entropy. Despite the large uncertainty associated to the temperature profile, and therefore to the total mass, the gas fraction
(red lines in Fig. 2) is found to be greater than the universal value at the virial radius. This may suggest the presence of a nonthermal pressure component, required to sustain HE in the cluster outskirts of Abell 2142. The point is strengthened by looking at the electron thermal pressure profile in absence of nonthermal pressure (red lines in Fig. 3) that is marginally consistent at $r \gtrsim r_{500}$ with that obtained by T16 deprojecting the SZ data via two different methods (with consistent results).

The black line in Fig.1 (top panel) shows the SM fit to the \textsl{XMM-Newton} temperature data when a nonthermal pressure component is included (see Eq. [A1]), and again a better fit is obtained for the bending entropy profile. The best fit parameter $a\approx 1.1$ is found for the pure powerlaw entropy profile; furthermore, for the bending entropy profile $r_b\approx 0.17\, R$ and $a'\approx 0.2$ are found, in turn yielding an entropy slope flattening from $a\approx 1.1$ in the inner regions toward the value $a_R\approx 0.12$ at the virial radius. The resulting deprojected temperature profile (see Fig. 1, bottom panel) implies a greater total mass after Eq. (A4), yielding a gas fraction in agreement with the universal value (see Fig. 2). This is obtained with $\delta_R \approx 0.5$ in Eq. (1), while the value of $l \approx 0.4$ is determined by looking for the best agreement of the total mass profile with the measurements reported in the literature (see Fig. 4, top panel).

The presence of a nonthermal component also implies that the thermal pressure profile is altered somewhat with respect to that in absence of nonthermal contribution. This is because in the HE equation ${\rm d}p_{\rm th}/{\rm d}r+{\rm d}p_{\rm nth}/{\rm d}r \propto - G\, M(<r)\,n/r^2$ the thermal and nonthermal pressure gradients cooperate to sustain the same gravitational pull; then it is easily understood that the solution of the HE equation for $p_{\rm th}(r)$ is necessarily different when a nonthermal pressure component is added.

The net outcome is that the electron thermal pressure $p_{\rm th}$ profile (black curve in Fig. 3) is now consistent out to the virial radius with that inferred from SZ data. This is a consequence of the behavior of $T$ toward the virial radius $R$ for $\delta_R>0$. Specifically, from the temperature and thermal pressure gradients derived by Eqs. (A1) and (A2) we get (Fusco-Femiano \& Lapi 2014)
\begin{eqnarray}
g_T &\equiv& (\frac{dlnT }{dln r})_R = \frac{3}{5}a_R-\frac{2b_R}{5(1+\delta_R)}\\
\nonumber\\
g_{p_{\rm th}} &\equiv& (\frac{dlnP }{dln r})_R = -\frac{b_R}{1+\delta_R}
\end{eqnarray}
where $b_R = (45-19a_R)/9$ (Cavaliere et al. 2009) and $a_R = a-(R/r_b -1)a^{\prime}$ (Lapi et al. 2010).

Thus for $\delta_R>0$ the SM temperature and thermal pressure profiles become flatter going toward the virial radius, in agreement with the measured SZ/X-ray profiles. We remark that the behavior of $T$ and $p_{\rm th}$ in
the cluster outskirts weakly depends on the adopted functional shape for $\delta(r)$ suggested by numerical simulations and expressed by Eq. (1), and that the same level of $p_{\rm nth}$ is able to satisfy both the universal gas fraction value and the SZ pressure profile observed by \textsl{Planck}. Recently, Ghirardini et al. (2017) have conducted an analysis on Abell 2319 adopting our pressure profile and showing its consistency with the observed SZ-inferred pressure profile.

As a final remark, we recall from Fusco-Femiano \& Lapi (2014) that, in the context of a model based on HE (possibly including a non-thermal component) like our SM, the detailed shape of the outer entropy profile
is more sensitively inferred from measurements of the temperature rather than of the pressure gradient. For Abell 2142, after Eqs. (2) and (3) it is seen that comparing a pure powerlaw entropy profile with $a=a_R\approx 1.1$ to a flattening entropy profile with $a_R\approx 0.12$ (corresponding to the best fit parameter $r_b/R\approx 0.17$ and $a'\approx 0.2$), yields a relative change in the thermal pressure gradient of $\Delta_P\approx 1-g_P(a_R=0.12)/g_P(a_R=1.1)\approx -0.8$ much smaller than in the temperature gradient $\Delta_T\approx 1-g_T(a_R=0.12)/g_T(a_R=1.1)\approx -3.4$.

\begin{figure}
    \includegraphics[width=\columnwidth]{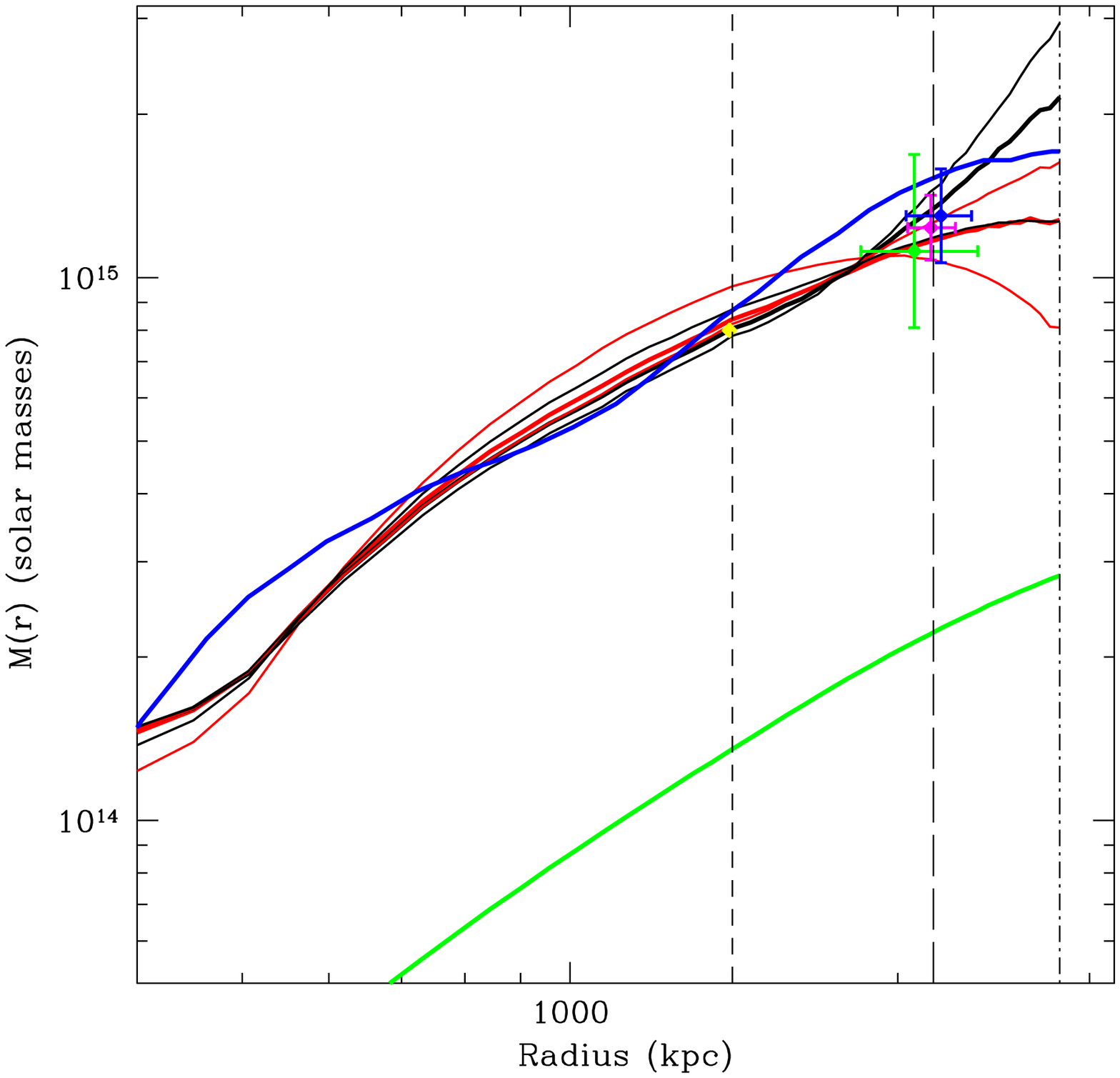}\\\includegraphics[width=\columnwidth]{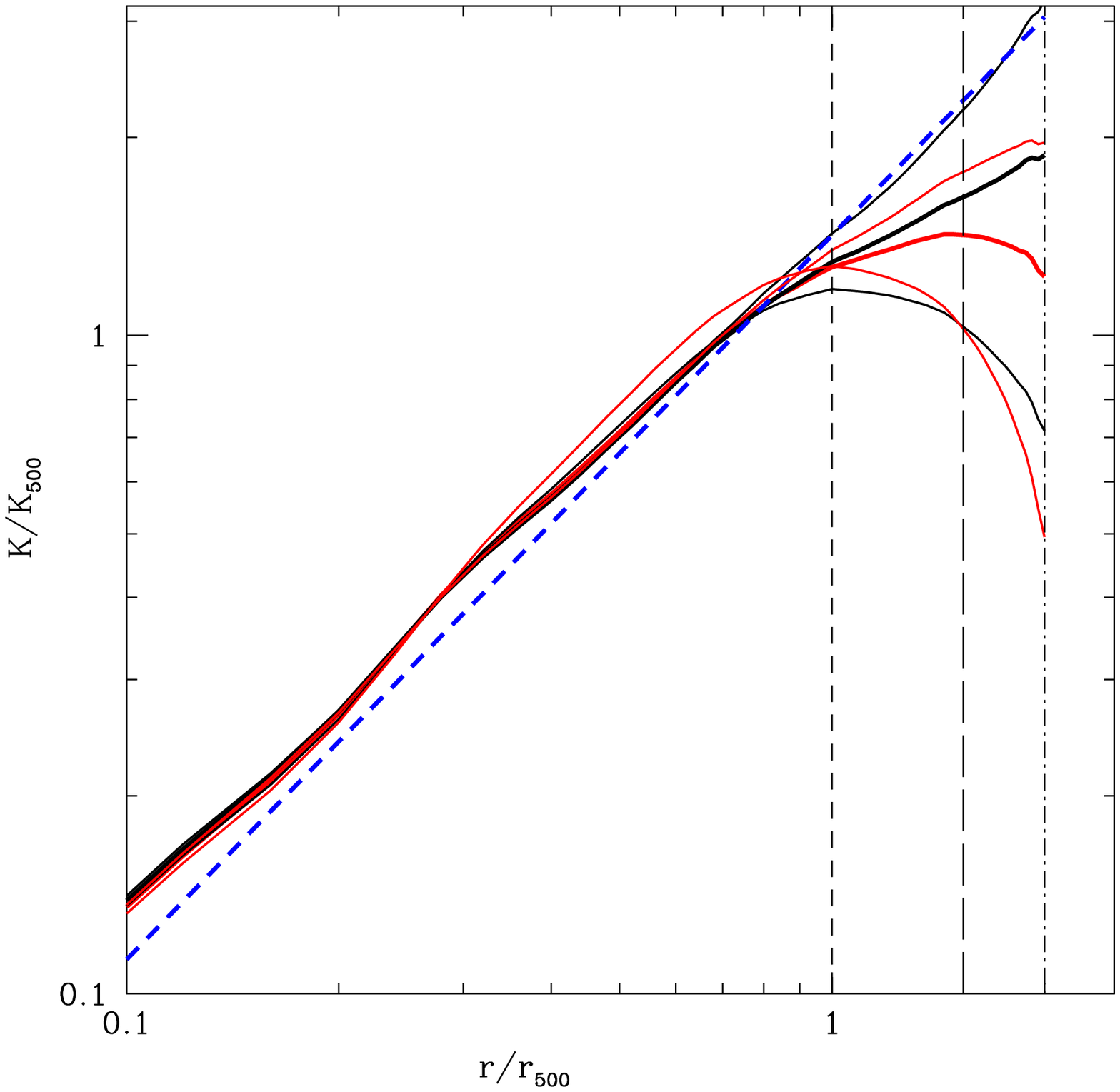}
    \caption{Abell 2142 - Top panel: Mass profiles. The red and black profiles represent the total mass obtained with $\delta_R = 0$ and $\delta_R = 0.5$ ($l$ = 0.4), respectively. The blue line is the central profile derived by the SZ/X-ray analysis (see T16). The green profile is the gas mass derived with the gas density corrected for clumpiness (see T16). For comparison, the yellow square is $M_{500}$ from $L_X - M$ relation (Piffaretti et al. 2011), green square is $M_{200}$ from \textsl{Suzaku} X-ray observations (Akamatsu et al. 2011), magenta square is $M_{200}$ from Subaru weak lensing (Umetsu et al. 2009), and black square is $M_{200}$ from galaxy kinematics (Munari et al. 2014). Bottom panel: Entropy profiles. Red and black lines are the entropy profile obtained via the deprojected temperature profiles of Fig. 1 (bottom panel) with $\delta_R = 0$ and $\delta_R = 0.5$ ($l$ = 0.4), respectively, and the gas density profile corrected for clumpiness (see T16). Blue dashed line shows the entropy profile expected from pure gravitational infall with slope 1.1 (Voit et al. 2005); for $r_{500}$ and $K_{500}$ see T16. The vertical dashed, long-dashed and dot-dashed lines represent $r_{500}$, $r_{200}$ and the virial radius $R$, respectively.}
\end{figure}

\section{Discussion and Conclusions}

High precision cosmology with galaxy clusters requires an accurate determination of cluster masses for wide sample of objects. However, estimates based on X-ray, SZ, and gravitational lensing observations differ substantially. A plausible explanation for this measurement bias involves
deviations from spherical symmetry and/or uncorrect description of nonthermal processes in cluster outskirts. Recently, several methods have been developed to constrain the nonthermal pressure support and hence to improve mass determinations based on the assumption of HE (Vazza et al. 2012; Shi \& Komatsu 2014; Shi et al. 2015, 2016; Biffi et al. 2016; Martizzi \& Agrusa 2016). In particular, the last authors show that nonthermal pressure support, though negligible in the inner regions, is increasingly relevant toward the outskirts, yielding a contribution of $(20-40)\%$ to the total pressure at radii in the range $r\sim r_{500} - 2\, r_{500}$. This contribution is considered a lower limit to the pressure balance in clusters because additional sources as magnetic fields and cosmic rays are currently not included in the simulations. The conclusion is that the total mass derived solely by the thermal pressure underestimates the actual cluster mass at least by $(10-20)\%$ at $r \gtrsim 0.5\,r_{500}$.

In the present paper we have performed via our SuperModel (SM) an analysis of the temperature data for Abell 2142 measured out to $r\lesssim r_{500}$ by\textsl{XMM-Newton}. Observations out to the virial radius will be fundamental for a more accurate determination of the ICM thermal state. Abell 2142 has been observed by \textsl{Suzaku} (Akamatsu et al. 2001) out to $r_{200}$ only in the NW direction of the possible merger axis. These temperatures are higher than the azimuthally averaged temperatures
measured by \textsl{XMM-Newton} in the study presented by T16. However, the SM fits to the \textsl{XMM-Newton} temperature data yield a total mass profile (see Fig. 4, top panel) consistent with the measurements reported in the literature at $r_{500}$ and $r_{200}$, showing the reliability of our analysis also at distances greater than $r_{500}$.  Another caveat is that the \textsl{XMM-Newton} temperature data are not corrected for the effects of clumpiness that may bias low the temperature profile. However, as it is claimed by T16, the effect of the presence of clumps on the spectroscopic temperature values is negligible.

The SM fit to the temperature profile requires an entropy run that flattens outwards for $r \gtrsim r_b\approx 0.17\, R$. By combining the deprojected temperature profile with the density profile corrected for clumpiness we have derived the gas mass $M_{\rm gas}$ and the total mass $M$ (Eq. [A4] for $\delta_R = 0$). The resulting gas fraction $f_{\rm gas} = M_{\rm gas}/M$ is greater than the universal value (see Fig. 2). Moreover, the thermal pressure profile derived by our SM analysis is in good agreement at $r \lesssim r_{500}$ with the profile obtained by deprojecting the SZ data, but deviates at larger radii (see Fig. 3). These results indicate the presence of a nonthermal pressure component required to sustain gravity in the cluster outskirts.

We have found that this additional nonthermal pressure component amounts to $\sim 30\%$ of the total pressure at the virial radius; such values are consistent with the indications from the empirical models and numerical simulations by Martizzi \& Agrusa (2016) and Nelson et al. (2014). We have shown that inserting a nonthermal pressure component of such an amount in the HE equation brings the gas fraction into good agreement with the universal value, because a greater value of $M$ is obtained for $\delta_R > 0$. The need to consider the presence of a nonthermal support in the cluster outskirts appears also from the $f_{gas}$ profile determined in T16 using the X-ray spectroscopic data that gives a gas fraction at $R$  greater than the universal value.

In addition, we have demonstrated that the presence of a nonthermal pressure component in the HE equation necessarily modifies the thermal pressure profile with respect to the behavior when such nonthermal component is absent. The net outcome is that the SM electron thermal pressure profile becomes flatter (see Eqs. 2 and 3) and consistent with the SZ data also for $r \gtrsim r_{500}$ (see Section 2). The corresponding SM temperature profile also becomes flatter, yielding a greater total mass at $R$ (see Fig. 4) and therefore a lower gas fraction consistent with the universal value. Note that in the central regions the nonthermal pressure support is negligible (cf. Fig. 3) so that the thermal pressure profile and the other thermodynamic quantities are unaffected.

We stress that the resulting total mass profile is consistent with the measurements reported in the literature at $r_{500}$ and $r_{200}$ (see Fig. 4, top panel). This shows the reliability of our SM analysis based on the fit to the \textsl{XMM-Newton} temperature data measured at $r\lesssim r_{500}$. The total mass at the virial radius is $2.15^{+0.79}_{-0.88}\times 10^{15} M_\odot$. Note that for deriving $M_{\rm gas}$ we consider the same gas density profile corrected for clumpiness inferred by T16; a slightly difference of $\sim 6\%$ is between our $M_{\rm gas}(<R)$ value and that reported in T16. Neglecting the nonthermal pressure biases low the total mass of $\sim 15\%$ at $r_{200}$ consistent with the values derived by numerical simulations (e.g., Martizzi \& Agrusa 2016).

We also stress that the method by T16 aimed at determining the temperature profile $k_B T_{SZ}=P_e/n_e$ from the SZ thermal pressure $P_e$ and the X-ray density $n_e$ does not allow to highlight the presence of a nonthermal pressure support in the cluster outskirts. Our SM analysis shows that when a nonthermal component is included in the HE, the thermal pressure profile is also affected; specifically, it is flattened outwards and becomes consistent with the observed SZ-inferred thermal pressure profile. This implies that the approach by T16, based on the SZ-inferred thermal pressure data, cannot highlight the presence of a nonthermal support. Eckert et al. (2017a), basing instead on X-ray \textsl{Chandra} observations of Abell 2142, recently report that infall of groups can generate turbulence in the cluster outskirts, consistently with our findings.

T16 also point out that the entropy flattening observed by \textsl{Suzaku} in the outskirts of several clusters can be explained by clumping with no need to invoke additional physics, and that the level of a nonthermal component, if present, may be negligible. We remark that these conclusions are not fully supported by the clumping factor $C = \langle n_e^2\rangle/\langle n_e\rangle^2$ (Mathiesen et al. 1999) measured in Abell 2142 by T16. They report at $r_{200}$ a clumping factor $\sqrt C = 1.18\pm 0.06$ that raises to $1.36\pm 0.13$ at the virial radius. These values are obtained removing clumps whose sizes are greater than the size of the Voronoi bins which corresponds to $\sim$ 20 kpc. Their measurements are in good agreement with simulations that include additional baryonic physics (Nagai \& Lau 2011: Roncarelli et al. 2013). However, the last value is insufficient to produce at $R$ an entropy level in agreement with the self-similar prediction. To satisfy this constraint $\sqrt C$ should be $\sim$ 1.7. This value is determined at the virial radius by the ratio between the entropy value given by the power law with slope $1.1$ and the value of $k$ obtained using the gas density profile affected by clumpiness ($C^{5/6}\approx 2.4$ for $k = P/n_e^{5/3}$, see their Fig. 12). The clumping factors indicated by simulations, or measured in real clusters like Abell 2142 (see T16), appear insufficient to explain the deviation from pure gravitational infall, as it is also the case for several other clusters (e.g., Fusco-Femiano \& Lapi 2014). This implies that the entropy deviation is partly due to a reduced entropy production and not exclusively to the presence of clumpiness in the gas density profile.

The nonthermal support highlighted by our SM analysis of Abell 2142 may arise by the increase of the bulk energy seeping through the virial boundary and driving turbulence in the outskirts, as suggested by the SM analysis of several other clusters; this occurs during the late cluster growth when the accretion shocks weaken (see Lapi et al. 2010; Cavaliere et al. 2011; Fusco-Femiano \& Lapi 2014). This mechanism accounts for the observed entropy flattening and for azimuthal variations in the cluster thermodynamic properties. A decreasing thermalization is more pronounced in cluster sectors adjacent to low-density regions of the surrounding environment or in the undisturbed directions of the cluster outskirts.

In conclusion, the SM analysis of the \textsl{XMM-Newton} temperature profile in Abell 2142 out to $r_{500}$ reveals the presence of a nonthermal pressure component at levels around $\sim 30\%$ of the total support in the cluster outskirts; this, jointly with a gas density corrected for clumpiness, yields at $R$ the universal gas fraction and the SZ pressure profile observed by the \textsl{Planck} mission. This nonthermal pressure component is not highlighted by determining the gas temperature from the SZ pressure data, as discussed above. On the other hand, the clumping factor profile may be derived by the X-ray brightness distribution, as reported by Tchernin et al. (2016) for Abell 2142, and the nonthermal pressure level by $f_{\rm gas}$, as shown by our SM analysis of the X-ray temperature. Another possibility to determine these two quantities is given by gas fraction measurements based on X-ray and weak gravitational lensing observations out to the virial radius. A X-ray mass lower than the lensing estimate establishes the level of the nonthermal pressure; relatedly, a gas fraction derived via the lensing mass that is found higher than the universal value indicates the degree of clumpiness in the gas density profile.

To obtain a more accurate determination of the nonthermal pressure level and of the degree of clumpiness requires X-ray observations out to the virial radius for several clusters. In the near future, a promising possibility is constituted by long-exposure \textsl{Suzaku} or \textsl{XMM-Newton} observations of cluster outskirts (e.g., the X-COP project; see Eckert et al. 2017b), although the sizes of the related samples will be limited to a few tens of clusters. In a future perspective, larger samples will likely become available thanks to the next generation of high-sensitivity X-ray observatories such as \textsl{Athena}.

\section*{Acknowledgements}
We are grateful to S. Ettori and R. Valdarnini for stimulating discussions. We thank the referee for valuable comments. AL acknowledges partial support by PRIN MIUR 2015 `Cosmology and Fundamental Physics: illuminating the Dark Universe with Euclid', PRIN INAF 2014 `Probing the AGN/galaxy co-evolution through ultra-deep and ultra-high-resolution radio surveys', and the RADIOFOREGROUNDS grant (COMPET-05-2015, agreement number 687312) of the European Union Horizon 2020 research and innovation programme.

\section{Appendix A}

In the presence of turbulence the total pressure is given by the thermal component and by an
additional nonthermal component. It can be written as
$P(r) = p_{\rm th}(r) + p_{\rm nth}(r) = p_{\rm th}(r)[1 +
\delta(r)]$ in terms of the quantity $\delta(r) \equiv p_{\rm nth}/p_{\rm
th}$ that when inserted in the HE equation yields the temperature profile as
\begin{eqnarray}
\nonumber\frac{T(r)}{T_R} &=& \left[\frac{k(r)}{k_R}\right]^{3/5}\, \left[\frac{1 + \delta_R}{1 + \delta(r)}\right]^{2/5}\, \left\{1 + \frac{2}{5}\frac{b_R}{1 + \delta_R}\right. \times\\
\\
\nonumber &\times & \left.\int_r^R {\frac{{\rm d}x}{x} \frac{v^2_c(x)}{v^2_R}\,
\left[\frac{k_R}{k(x)}\right]^{3/5}\, \left[\frac{1 + \delta_R}{1 +
\delta(x)}\right]^{3/5}}\right\}
\end{eqnarray}
and the thermal pressure profile as
\begin{eqnarray}
\nonumber \frac{P(r)}{P_R} &=& \left[\frac{1 + \delta_R}{1 + \delta(r)}\right]\left\{1 + \frac{2}{5} \frac{b_R}{1 + \delta_R}\right. \times\\
\\
\nonumber &\times& \left.\int_r^R {\frac{{\rm d}x}{x}
\frac{v^2_c(x)}{v^2_R} \left[\frac{k_R}{k(x)}\right]^{3/5}\, \left[\frac{1 +
\delta_R}{1 + \delta(x)}\right]^{3/5}}\right\}^{5/2}
\end{eqnarray}
where $v_c$ is the circular velocity ($v_R$ is the value at the virial radius $R$), and $b_R$
is the ratio at $R$ of $v^2_c$ to the sound speed squared (Cavaliere et al. 2009; Cavaliere et al. 2011).

For $k(r)$ we consider the spherically averaged profile with shape
$k(r) = k_c + (k_R - k_c)(r/R)^a$ (see Voit 2005). To model the reduced entropy production relative to a pure gravitational inflow, reported by several X-ray observations, the SM considers an entropy run that starts as a
simple powerlaw with slope $a$, but for radii $r \gtrsim r_b$ deviates downward (Lapi et al. 2010). For the sake of simplicity, the entropy slope is taken to decline linearly with a gradient $a^{\prime} \equiv (a-a_R)/(R/r_b - 1)$, where $r_b$ and $a^{\prime}$ are free parameters to be determined from the fitting of the X-ray observables.

The weakening of the accretion shocks involves more
bulk energy to flow through the cluster, and drive turbulence into the
outskirts (Cavaliere et al. 2011b). Turbulent motions originate at the virial
boundary and then they fragment downstream into a dispersive
cascade to sizes $l$. Numerical simulations show that small values of the
turbulent energy apply in the cores of relaxed clusters, but the ratio
$E_{\rm turb}/E_{\rm thermal}$ of the turbulent to thermal energy increases
into the outskirts (e.g., Vazza et al. 2011). For $\delta(r)$ we use the functional shape
\begin{equation}
\delta(r) = \delta_R\, e^{-(R-r)^2/l^2}
\end{equation}
which decays on the scale $l$ toward the internal cluster regions from a round maximum at $R$. This profile of
$\delta(r)$ concur with the indication of numerical simulations (Lau et al.
2009; Vazza et al. 2011).

The traditional equation to estimate the total X-ray mass $M(r)$ within $r$
is modified as follows to take into account the additional nonthermal
pressure component (Fusco-Femiano \& Lapi 2013)
\begin{eqnarray}
\nonumber M(r) &=& - \frac{k_B [T(r)(1 +\delta(r)] r^2 }{\mu m_p G}\left\{
\frac{1}{n_e(r)}\frac{d n_e(r)}{d r}+\right. + \\
\nonumber\\
\nonumber&+& \left.\frac{1}{T(r)[(1+\delta(r))]}\frac{d T(r)[1 +
\delta(r)]}{d r}\right\}=\\
\\
\nonumber&=& - \frac{k_B [T(r)(1 +\delta(r)] r^2}{\mu m_p G}\left[\frac{1}{n_e(r)}
\frac{d n_e(r)}{d r} \right. + \\
\nonumber\\
\nonumber &+& \left. \frac{1}{T(r)}\frac{d T(r)}{d r}+\frac{\delta(r)}{1 + \delta(r)} \frac{2}{l^2}(R - r)\right]~.
\end{eqnarray}
The hot gas mass writes
$M_{\rm gas} = 4\pi \mu_e m_p\int{\rm d}r~{n_e(r) r^2}$
where $\mu_e \sim 1.16$ is the mean molecular weight of the electrons.

\bsp
\label{lastpage}
\end{document}